\documentclass[conference]{IEEEtran}
\usepackage{cite}

\ifCLASSINFOpdf
  \usepackage[pdftex]{graphicx}
  \usepackage[outdir=./]{epstopdf}
  
  \graphicspath{{figures/}}
  \DeclareGraphicsExtensions{.pdf,.jpeg,.png,.eps}
\else
  \usepackage[dvips]{graphicx}
  \graphicspath{{figures/}}
  \DeclareGraphicsExtensions{.eps,.jpg,.png}
\fi
\begin{document}
	
\newcommand\copyrighttext{ 
	\Huge {IEEE Copyright Notice} \\ \\
	\large {Copyright (c) 2019 IEEE \\
		Personal use of this material is permitted. Permission from IEEE must be obtained for all other uses, in any current or future media, including reprinting/republishing this material for advertising or promotional purposes, creating new collective works, for resale or redistribution to servers or lists, or reuse of any copyrighted component of this work in other works.} \\ \\
	
	{\Large Accepted to be Published in: Proceedings of the 2019 IEEE International Conference on Industrial Cyber-Physical Systems (IEEE ICPS 2019), May 06-09, 2019} \\ \\ 
}

\twocolumn[
\begin{@twocolumnfalse}
	\copyrighttext
\end{@twocolumnfalse}
]
	
%
\title{MAIA: A Microservices-based Architecture\\for Industrial Data Analytics}

\author{\IEEEauthorblockN{Hai Dinh-Tuan}
\IEEEauthorblockA{Service-centric Networking\\Telekom Innovation Laboratories\\Technische Universit{\"a}t Berlin\\
hai.dinhtuan@tu-berlin.de}
\and
\IEEEauthorblockN{Felix Beierle}
\IEEEauthorblockA{Service-centric Networking\\Telekom Innovation Laboratories\\Technische Universit{\"a}t Berlin\\
beierle@tu-berlin.de}
\and
\IEEEauthorblockN{Sandro Rodriguez Garzon}
\IEEEauthorblockA{Service-centric Networking\\Telekom Innovation Laboratories\\Technische Universit{\"a}t Berlin\\
sandro.rodriguezgarzon@tu-berlin.de}}




\maketitle

\begin{abstract}
In recent decades, it has become a significant tendency for industrial manufacturers to adopt decentralization as a new manufacturing paradigm. This enables more efficient operations and facilitates the shift from mass to customized production. At the same time, advances in data analytics give more insights into the production lines, thus improving its overall productivity. The primary objective of this paper is to apply a decentralized architecture to address new challenges in industrial analytics. The main contributions of this work are therefore two-fold: (1) an assessment of the microservices' feasibility in industrial environments, and (2) a microservices-based architecture for industrial data analytics. Also, a  prototype has been developed, analyzed, and evaluated, to provide further practical insights. Initial evaluation results of this prototype underpin the adoption of microservices in industrial analytics with less than 20ms end-to-end processing latency for predicting movement paths for 100 autonomous robots on a commodity hardware server. However, it also identifies several drawbacks of the approach, which is, among others, the complexity in structure, leading to higher resource consumption.

\end{abstract}



%
\IEEEpeerreviewmaketitle

\section{Introduction}

Current manufacturing systems are still mostly based on the 5-layer architecture (ISA-95 model) \cite{isa2000isa}, which was originally developed to ease the management of interfaces between enterprise application systems and manufacturing controllers. However, the top-down decision making process in this model is no longer suitable for future factories, in which the concept of \textit{Fog Computing} pushes more decisions to be made at the lower levels of the infrastructure. At the same time, the fierce competition among manufacturers force them to build more agile and responsive production systems. These factors have led to many projects worldwide to develop new manufacturing techniques, focusing on four main strategies, as Franciso \cite{almada2016industry} summarized: \textit{Decentralization} (both physically and logically); \textit{Vertical integration, connectivity and mobile}; \textit{Cloud Computing}; and \textit{Advanced analysis}. 

\begin{figure}
	\centering
	\includegraphics[width=0.9\linewidth]{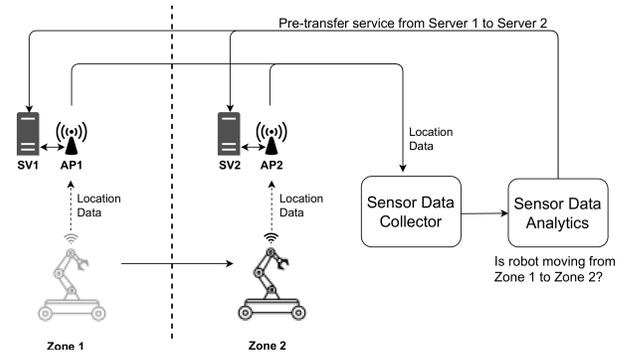}
	\caption{The use case of predictive analytics for low-latency fog computing used to evaluate our architecture. As an autonomous robot moves from Zone 1 to Zone 2, the proposed system is expected to be able to predict this movement and notify the fog computing infrastructure.}
	\label{fig:boat1}
\end{figure}

This work contributes to such strategies by evaluating a decentralized software architecture for industrial analytics. Together with our industry partners, we have identified a use case as depicted in Fig.1, illustrating a manufacturing factory with multiple \textit{zones}. In each \textit{zone}, we set up a fog computing infrastructure composed of one access point (AP) and one server (SV) in close proximity. An autonomous robot connects to the closest access point and, due to the limitation of computing resources and battery life, off-loads the object recognition tasks to the server by streaming the video captured from its camera. To avoid collisions with other robots or objects in the shop floor, the processing latency must be always kept as low as possible. However, as the robot moves, it might connect to other access points along the way and the latency increases (in this scenario, the robot is moving from \textit{Zone 1} to \textit{Zone 2}, thus switching from \textit{AP1} to \textit{AP2}). The low-latency requirement is guaranteed only when the processing service will be dynamically moved to another server that is closer to the robot's new location, i.e. \textit{SV2}, with minimal interruption. Therefore, we propose setting up a microservice-based sensor data analytics component to predict robot's movement path and the next access point it will connect to by analyzing location data from the robots in real-time. These predictions are then transferred back to the fog computing infrastructure, thus allowing parts of the service to be transferred to the new location in advance and reducing the overall interruption.

The rest of this paper is organized as follow. In Section II, we identify three major trends in industrial analytics and how microservices fit into that developments. Various efforts to apply microservices in industry in general and in manufacturing in particular are reviewed in Section III. Section IV describes the design criteria, design decisions, and our proposed architecture in detail. More about the implementation of our prototype are given in Section V. The evaluation of this prototype and the key results are discussed in Section VI. We conclude the findings and suggest some directions for future research in Section VII.

\section{Research background}

Recently, advanced data analytics have been increasingly applied across industries, covering the whole process of extracting useful knowledge, discovering patterns, and predicting trends from the collected data \cite{hromic2015real}. Analytics, especially cloud-based analytics will play a more significant role in the near future, driven by three important trends.

First, alongside the increasing adoption of \textit{Industrial Internet of Things (IoT)}, the number of sensors, and its measured data will continue to multiply in volume, velocity, and complexity \cite{lee2015cyber}. On one hand, it will be a valuable source of data, providing even more insights into underlying processes. On the other hand, however, this huge amount of information poses significant challenges for analytical systems, especially when the time constraints are taken into account.

Since more data has already been made available, there is a shift of focus from data collection toward data analytics. Data analytics in industry can transform data into information, knowledge, and insights, enabling multiple forms of data-driven intelligence. However, the heterogeneous nature of sensor data, the highly distributed data sources, and the security requirements raise several challenging issues for the computing platform. New researches are required to make current infrastructures more scalable, flexible, and robust. 

Third, as the data analytics techniques become more complex, more computing power is required, motivating manufacturers to adopt cloud technologies. This technology enables a number of innovative manufacturing models, such as distributed small-scale local manufacturing, loosely coupled manufacturing ecosystems, and agile manufacturing, paving the way for mass customization, which is identified as the new nature of consumer demand \cite{vincent2018future}. Again, the adoption of cloud computing raises challenges in guaranteeing the \textit{Quality of Service}, data security and privacy, etc.

At the same time, microservices has gained attractions from both the industry and academia due to the adoption at multimillion users companies such as Amazon or Netflix. According to \cite{lewis2014microservices}, a microservices-based application is a single application, composed of a set of services. These services are not only small in size but also highly decoupled, independently replaceable, upgradeable, and deployable \cite{krylovskiy2015designing}, designed following the well-known UNIX principle "Do one thing and do it well" \cite{newman2015building}. Despite being highly independent, i.e., each service has its own process and data management mechanism, there is an extensive communication flow between them. These interactions are not done using internal calls, but rather through clearly defined interfaces, such as HTTP-based API. Similarly, various types of clients can interact with a microservice-based application using protocols such as REST.

Our hypothesis is, the distributed, and cloud-native nature of microservices make it a potential approach for industrial analytics. First, by decomposing an application into several smaller services, microservices can achieve \textit{fine-grained scalability}, allowing certain services to be scaled up/down while keeping other services intact. This means, compared to traditional monoliths, microservices can be easily scaled in different deployment strategies (in both cloud and edge networks), providing manufacturers more flexibility to optimize the latency, energy, and resource utilization. 

Also, by decomposing a software into multiple smaller deployment units, the interdependence among microservices is minimized. Each microservice can be deployed or updated without redeploying the whole application. For example, new components processing new sensor data types with different analytics techniques can be added into an existing system without any interruption. Also in case several components break, the isolation among services can help preventing cascading failures. Therefore, the microservices architecture improves the availability as well as the robustness of critical systems in manufacturing environments. 

Third, the autonomy of microservices allows the most appropriate technology to be applied where it can solve problems most effectively. This capability is also referred to as \textit{technology heterogeneity} or \textit{polyglot programming and persistence}, allowing different technologies to coexist in an application. Not only the manufacturer can maximize the efficiency of each individual process, but also the developers can select the tool-set that they are most familiar with to improve the quality and the speed of service deployment.

From a business and management perspective, microservices-based software can be delivered continuously using agile development methodologies and cycles, allowing manufacturers to react quickly to changing customers and market needs. Small- and medium-sized factories also benefit from this architecture as it enables them to leverage external expertise and resources through outsourcing without revealing the core business functions \cite{rossberg2012pro}. This feature realizes the vision of smaller factories located closer to the customer, thus be able to quickly deliver individualized products that meet customers' unique requirements.

On the other hand, we should take into consideration several drawbacks of this concept. First of all, as a distributed model, microservices introduce its own set of complexities, for example, additional components to orchestrate and monitor services are required. Developers also have to face the typical problems of a distributed system such as data management or complex communication patterns. Moreover, more components needs to be implemented and deployed exposes more potential failure points, requiring sophisticated measures to isolate the failures and automatically restore failed components.

In addition, an application with multiple independent microservices is expected to consume more hardware resources than a single monolith. That is because each microservice runs in a separated process or thread, or even in a separate virtual machine (VM) such as Java virtual machine (JVM). Today, a common practice to fasten the deployment process and improve the independence on underlying platform is to adopt container or virtualization technologies. However, this strategy requires even more hardware resources. 

\section{Related work}

Ciavotta et al. \cite{ciavotta2017microservice} propose a simulation-based architecture for \textit{Cyber-Physical Systems (CPSs})\footnote{Cyber-Physical systems are systems with the seamless, real-time interaction between computing elements and physical assets using intelligent data management, analytics and computational capability \cite{lee2015cyber}.} at shop-floor level, providing an environment for \textit{Digital Twins (DTs)}\footnote{Digital Twins are representations of real-world assets (including the assets in designing/building stage) created with the ability to collect and synthesize data from various sources including physical data, manufacturing data, operational data and insights from analytics software \cite{grieves2014digital}.} along the whole plant life-cycle. The proposed platform implements a microservice IoT-Big Data architecture supporting the publication of multidisciplinary simulation models and managing streams of data coming from the shop-floor for real-digital synchronization. Microservices architecture is applied in their support infrastructure, in order to manage the DTs. In our proposal, we extend this work by employing microservices also for simulating physical assets to decentralize the whole system. Rather than storing all digital copies of physical assets in one place, building them as microservices allows more flexibility in deployment strategies. For each service, a best physical location to deploy could be determined based on various criteria. This is an important foundation to enable automated and QoS-aware deployment strategies.

Thramboulidis et al. \cite{thramboulidis2018cyber} describe a 5-layer framework based on microservices for manufacturing systems. Each physical unit of the plant is transformed to a smart entity, named a \textit{Cyber-Physical Microservice} (CPMS). The authors define two main types of CPMS: \textit{Primitive CPMSs}, which encapsulates a physical artifact and transform it to a smart entity, and \textit{Composite CPMSs}, which utilizes at least one \textit{Primitive CPMS}. In this work, the authors focus on the architecture of each individual CPMS and evaluate the overhead of microservice orchestration. Our paper is complementary to this work, as it proposes a complete architecture for utilizing CPMSs.

NIMBLE Collaborative Platform \cite{innerbichler2017nimble} adopts microservices architecture to build a collaborative Industrie 4.0 platform that enables IoT-based real-time monitoring, optimization and negotiation in manufacturing supply chains. The microservices in the architecture provide beyond the core business functionalities essential supporting services such as Gateway Proxy, Service Logging, Service Monitoring, Service Discovery, Service Configuration, Identity Management. The implementation uses either REST (i.e. HTTP) or messaging as the mean of service communication. The authors describe from an architectural viewpoint how they use microservices to build a platform processing incoming data but they don't address the scalability of the platform. In our proposed architecture, we incorporate the self-management capability for the system by introducing a new component named \textit{Life-cycle management}. Its main responsibility is to monitor the load level of each microservices, and scale up or down a specific microservice based on various criteria such as memory consumption, incoming message queue length, etc.

In \cite{vianden2014experience}, the authors propose a microservice-based reference architecture for the \textit{Enterprise Measurement Infrastructure (EMI)} . In this architecture, there are six main components, which are Visualization (dashboard tools to interact with users), Calculation and Storage (provides aggregated information for Visualization layer), Data Transport and Integration (builds a common communication infrastructure for all services), Data Provider (feeds data to the system), Data Adapter (convert received data into readable formats) and Operation (contains services that ease operating and monitoring the application). This architecture is a basic design, as it is built to exploit most prominent advantages of microservice: high level of isolation between services, robust again complete system failure, and the support for the integration of heterogeneous systems. However, this architecture again does not address the scalability of a microservice-based application.

The \textit{Reference Architecture Model Industrie 4.0 (RAMI 4.0)} \cite{hankel2015reference} is a three-dimensional model for service-oriented architecture, combining the life-cycles of products, factories, and machinery with the hierarchy levels of Industrie 4.0. The authors define six layers, from top to bottom are: Business, Functional, Information, Communication, Integration, and Asset. The Integration layer is the connection between the digital world (four upper layers) an the real world (the bottom layer). As RAMI is developed as a generic framework for industrial applications in general, a more concrete architecture designed for decentralized analytics will provide new perspectives by addressing specific challenges of the concept.

\section{Concept and design}

\subsection{Design criteria}

This work focuses on applying the decentralization paradigm in manufacturing as an approach to be flexible and able to react faster to market and customer demands. Indeed, in \cite{zhang2006analytical}, the authors have categorized four paradigms for manufacturing systems, all of them are only possible when the hardware and software components are reconfigurable, which encourages the adoption of modular structures and open architectures. Besides decentralization, we also consulted the \textit{Industrie 4.0 design principles} \cite{hermann2016design} in order to compile a list of additional design criteria as below:

\textbf{Scalability:} An industrial application should be flexible enough to scale in different dimensions: the number of nodes, functionalities,  applications, and the data volume size, among others. Unlike other networks, a network of industrial devices can scale up to millions of connected points. This leads to several architectural requirements, such as selective data transmission (ensure the data can only be sent to the necessary parts of the system) and a well-defined set of interfaces to support fast system integration.

\textbf{Low-latency capability:} Low-latency computation is an important requirement for industrial applications in general. Many applications are realized only with on-time decisions, for example, autonomous robots normally require latency as low as tens of microseconds \cite{schneider2017industrial}. In a distributed computing model, this requirement is challenging to fulfill, since the time required for processing might be affected by external factors, such as network conditions or the communication mechanisms between components \cite{hadlich2012time}.

\textbf{Interoperability:} One primary requirement for the design is a high level of interoperability, which is, among others, reflected in the use of open standard protocols or interfaces and multi-platform technologies such as HTTP or HTML. In addition, the application should be able to be deployed with ease on various run environments.

\textbf{Fault-tolerance:} Interruptions are unacceptable for critical mission software, therefore, the design must incorporate some recovery mechanisms after failures.

\textbf{Usability:} The application must provide intuitive methods for users to interact with. Monitoring capabilities must be available and provide useful operation's statistics in real-time.

\subsection{Design decisions}

Considering the previously mentioned requirements, our proposed MAIA architecture (\textbf{M}icroservices-based \textbf{A}rchitecture for \textbf{I}ndustrial data \textbf{A}nalytics) is designed based on six key design decisions:

The decentralized nature of the design is reflected in the adoption of microservices architectural style itself.  We define two main type of services: \textit{Functional services} and \textit{Infrastructure services}. \textit{Functional services} are services that support specific business operations or functions, whereas \textit{Infrastructure services} support nonfunctional tasks such as authentication, authorization, auditing, logging, and monitoring. This is an important distinction, because \textit{Infrastructure services} are not exposed to the outside world but rather are treated as private shared services only available internally to other services. In contrast, \textit{Functional services} provide their services externally.

We applied \textit{Domain-driven design} and the \textit{Bounded context} concept to define microservices with minimal inter-services dependency. In our design, each physical machine in the factory has a digital replica with all the related data and processes represented by a microservice, similar to the concept of CPMS proposed in \cite{thramboulidis2018cyber}. Although this design requires more hardware resources comparing to have a single service to handle multiple manufacturing units, we can tailor the service to meet the specifications of physical entities and better support the portability of these virtual representation. While simple analytics and monitoring tasks can be done in each individual digital representation of physical assets, the data is distributed across different digital replicas. To solve more complex problems such as optimization or predictive analysis, aggregated data from multiple machine is required. Therefore, we design two level of analytics, corresponding to two level of knowledge: individual analytics and global analytics.

The main mechanism for the inter-service interaction is \textit{asynchronous messaging} to decouple services. This means services do not communicate directly with each other, but rather by sending messages via a message broker. One advantage of this design is that the services don't need to wait for other to response to a request. Additionally, they don't need to know exact location of other service instances. However, this design leads to a single point of failure, which should be compensated by deploying the message broker in high availability clusters.

Inspired by the EMI reference architecture, we apply the database per service pattern. Together with the asynchronous messaging, this helps achieving a high level of autonomous and independence between microservices. This design also allows developers to incorporate some self-healing capabilities into their system, because in the event of failure, each microservice can independently restore its data, without affecting data from other services. However, a distributed data storage model would lead to some problems in ensuring data consistency across services. To solve this issue, we encourage developers to adopt the \textit{Eventual Consistency} and \textit{Event-driven model}. In this concept, whenever a microservice needs to update its database, it also publishes an event to an event message queue. By subscribing to that queue, other microservices are informed and can update their own databases accordingly.

To enable the fault-tolerance capability of the architecture, we incorporate two mechanisms. First, a monitor service is set up to check the health status of each deployed microservice and restart it if error occurs. On top of that, the \textit{Circuit Breaker} pattern is applied to specify a fallback method for microservices, thus gracefully degrading the functionalities of the application and avoiding cascading failures.

We deploy each microservice as a containerized unit, which can minimize the burden of incompatibility between platforms. Using container technology also adds an additional layer of management on top of our architecture, allowing the incorporation of third-party solutions for visualization, monitoring, and management. 

\subsection{MAIA Architecture}

The MAIA architecture consists of four main building blocks: \textit{Physical Space}, \textit{Digital-Physical Integration}, \textit{Digital Space}, and \textit{Infrastructure Services}. All components, except the \textit{Physical Space}, are composed of multiple microservices as shown in Fig. 2:

\begin{figure*}
	\centering
	\includegraphics[width=0.7\linewidth]{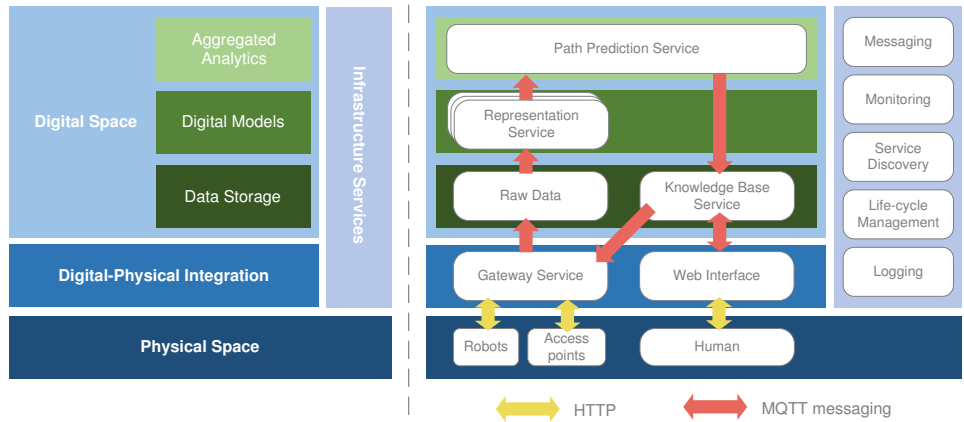}
	\caption{Our proposed Microservices-based Architecture for Industrial Data Analytics and the prototype implemented.}
	\label{fig:boat1}
\end{figure*}

\textbf{\textit{Physical Space:}} The first component in this building block is sensor-equipped machines in the shop floor. They are connected with upper components via distributed gateways using HTTP REST API or a messaging protocol. The interactions are done bidirectionally: The machines report various types of data and received commands, configurations adjustments, etc. from the analytics system in return. The human operators are also part of this MAIA component, however, they interact with the system via a web-based interface with reports, summary, visualized statistics in order to quickly gain insights into processes in the shop-floor. The operators can also change the configurations of machines directly via the same interface.

\textbf{\textit{Digital-Physical Integration:}} This is the interface for all the interactions between the physical and virtual world, consisting of two bidirectional interfaces: \textit{Web-based interface} for human interaction, providing visualizations, reports, summary, etc.; and \textit{API Gateway} provides physical entities a secured channel to exchange data with the system. With the \textit{API Gateway}, the design provides an abstraction level for the underlying connection technologies. Since various types of connections are installed in modern factories, from wired to wireless connections, an abstraction layer helps decoupling the architecture from physical connection technologies. Each driver for one connection technology can be implemented as a microservice, keeping other interactions unchanged. Other functions such as access control, authentication, and authorization are also implemented in this building block.

\textbf{\textit{Digital Space:}} This component stays at the heart of the entire architecture, providing representations of physical equipment and related processes in virtual world. There are three sub-components in this space, as explained below.

The \textit{Data Storage} provides a distributed big data storage infrastructure, with the ability to store a big amount of data, with high rates of random write and access, flexible enough to manage different data models, supporting both structured and unstructured data. Raw data collected from the physical systems is stored in the \textit{Raw Data} and distributed to the corresponding microservices. The knowledge extracted from analytics processes is synthesized in the \textit{Knowledge Base}, accessible for services in the \textit{Digital-Physical Integration}.

In \textit{Digital Models}, the virtual representations (Digital Twins) of the physical devices are built and modeled with all of their dynamics. To ensure their accuracy, these representations contain a model of the physical artifact combined with  its collected data. These digital models are structured in single component level and composed component level, similar to how various components compose a machine in physical world. Individual analytics, which can be done using data from one physical entity, are also performed here and their results are collected at the \textit{Knowledge Base}. Depending on the types of analytics, (requirements of processing power, processing time, and data privacy), this digital copy of assets can be deployed either at the network edge (low latency, context awareness), or in cloud infrastructure (slower but posses more computational resources). Simple management functionalities for digital models such as monitoring, inventory management, etc. are also included here.

Advanced analytics are performed in the \textit{Aggregated Analytics}. Such analytics require data from more than one physical entity, for example global optimization. Although these analytics require more time to perform, their outputs are essential for strategic decisions. Again, similar to short-term analytics made in each individual digital twin, the knowledge is communicated to the \textit{Knowledge Base}. 

\textbf{\textit{Infrastructure Services:}} While the three components mentioned above are designed to be compatible with the RAMI 4.0  \cite{hankel2015reference}, this component provides specific functionalities for a microservices-based application, including but not limited to service discovery, internal communication services, service logging, and monitoring. In addition, we also include the \textit{Life-cycle Management} capability as a component responsible for monitoring performance metrics such as CPU, memory load, etc. and make decision when to scale up or scale down a certain service by adding/removing service instances. 

\section{Implementation}

To evaluate the proposed architecture, we have implemented a Java-based prototype for the low-latency fog computing use case mentioned earlier.  The containerized microservices are categorized into \textit{Infrastructure services} and \textit{Functional services}. The \textit{Functional services} are implemented as below:  

\textbf{\textit{Gateway Service}}: Corresponding to the \textit{Digital-Physical Integration} in the proposed architecture, its main responsibility is to expose a REST API for robots and access points to update their data. Upon the receipt of an update, this service is responsible for (1) persisting this data, and (2) creating an event to notify other services using \textit{Eventual Consistency}.

\textbf{\textit{Representation Service}}: This service implements the \textit{Digital Models} component of the MAIA architecture. Each robot is modeled with a robot ID, current location in latitude/longitude coordinates and associated AP. During operation, multiple instances of this service run simultaneously, each processes the entire data collected from an individual robot and continuously monitors the distance from the robot to its connected AP. If this distance exceeds a pre-defined threshold, the service assumes this robot is leaving its current \textit{zone} and send a message to the \textit{Path Prediction Service}, triggering the prediction process.

\textbf{\textit{Path Prediction Service}}: This service implements a typical service in the \textit{Aggregated Analytics}. This service receives requests from instances of \textit{Representation Service} and performs a predictive analysis to predict the movement direction of the robot. Compared to individual instances of \textit{Representation Service}, this service has a global knowledge of APs' locations and coverage areas, which then used to correlate with the robot's predicted path to find out the next AP the robot may connect. This service outputs recommendations, which includes the robot ID and a list of maximum three possible APs ranked by the recommendation's confidence. 

\textbf{\textit{Knowledge Base Service}}: This service is responsible for keeping track of recommendations. It maintains two connections with (1) the \textit{Web-based Interface} to display recommendations received from the \textit{Path Prediction Service}, and  (2) the fog infrastructures to form a feedback loop for location updates.

\textbf{\textit{Web-based Interface}:} Provides a portal with visualized network of microservices, together with runtime metrics such as health status, memory, cache, system and environment properties, etc. The recommendations are also displayed here. 

The \textit{Infrastructure services} consist of the following services:

\textbf{\textit{Message Broker}:} In \cite{thramboulidis2018cyber}, the authors used Lightweight M2M based on Constrained Application Protocol (CoAP) as the main message transportation protocol. In contrast to this work, we propose to use Message Queue Telemetry Transport (MQTT) as the main protocol for microservices' interactions. Compared to CoAP, MQTT has more sophisticated reliability and congestion control mechanisms, which becomes significant when data is exchanged frequently \cite{de2013comparison}. In addition, when the packet loss rate is low, MQTT performs better in terms of latency \cite{thangavel2014performance}. However, it should be noted that CoAP, which is based on UDP, is more lightweight compared to a TCP-based protocol like MQTT. The \textit{Message Broker} manages several MQTT message queues: \textit{Representation service’s queue}: Each instance of the \textit{Representation Service} has a separate queue to receive updates from the \textit{Gateway Service}; \textit{Aggregation queue}: delivers messages from the \textit{Representation Service} to the \textit{Path Prediction Service}; \textit{Knowledge Based queue}: delivers recommendations from the \textit{Path Prediction Service} to the \textit{Knowledge Base Service}; \textit{\textit{Data event queue}}: delivers data update events created by the \textit{Gateway Service} to other services, thus allowing services to synchronized data about registered robots and access points.

\textbf{\textit{Service Registration and Discovery}:} This service allows microservices to register themselves, as well as discover other microservices dynamically during runtime. In addition, it also keeps track of deployed microservices and provide APIs for other management services. This service is implemented with two smaller components: (1) an entity receives registrations from microservices and (2) a REST client in each microservice that registers itself with the registry. To mitigate the single-point-of-failure risk, we deployed two instances of this registrar, continuously synchronizing  data with each other.

\textbf{\textit{Service Monitor and Management}:} We create a number of HTTP endpoints for each microservices to expose operational data such as health status or resource consumption level. A microservice is deployed to gather all these data to monitor the application and visualize on a interface. This service also restarts other deployed microservices in case of failure.

\textbf{\textit{Life-cycle Management}:} This service monitors the message queue length and automatically scale up the subscribing microservice if the number of unprocessed messages exceed a certain threshold. When the demand decreases, this service also reduces the number of service instances.

\textbf{\textit{Service Logging}:} The main responsibility of this service is to gather all application logs from other microservices into a single place. This service also provides useful statistics regarding the request propagation between multiple microservices.

\begin{figure}
	\includegraphics[width=\linewidth]{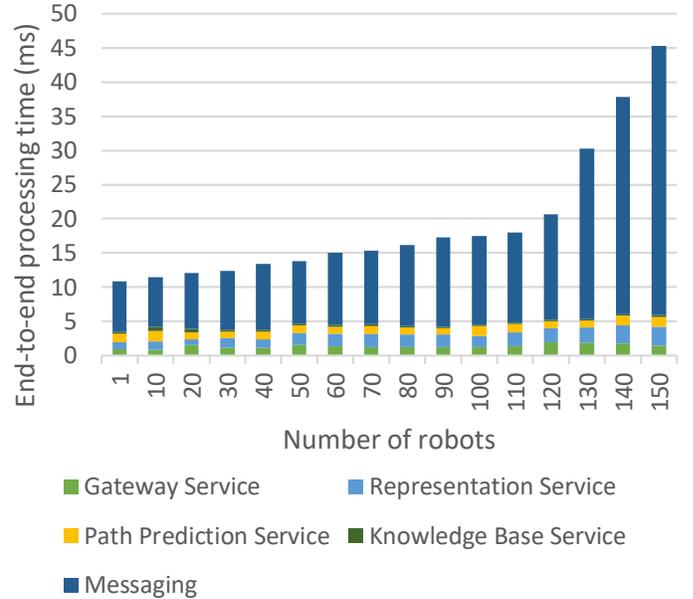}
	\caption{End-to-end processing time in different test cases.}
	\label{fig:boat1}
\end{figure}

\section{Empirical Evaluation}

\subsection{Results}

With the evaluation, we want to investigate the performance of microservices in terms of processing latency, as well as the overhead of containerization. The containerized prototype is deployed with Docker Compose in a dedicated, cloud-based server with commodity hardware (Intel i7-6700, 64GB DDR4, 4TB SATA HDD). Robots' location updates are sent to the server via an Internet connection with the frequency of 1Hz.

To evaluate the latency performance of the prototype, we perform \textit{Distributed Tracing} by adding a unique end-to-end identifier to each message, hence enable tracking the processing time of each microservice. 16 test cases with the number of robots in each case varies from 1 to 150 robots have been conducted.  In each case, the test was performed three times, and the mean values of processing time were recorded. We define the processing time of each microservice is the period between the arrival of a message until that message leaves. The end-to-end processing time for each request is the total of processing time in each microservice plus the time for message transportation between microservices, i.e. the period counted since the request arrival at the \textit{Gateway Service} until a recommendation is made and sent to the Web interface by the \textit{Knowledge Base Service}. 

The recorded results are presented in Fig. 3 with two key findings: (1) The end-to-end processing time increases from 10.872ms to 45.293ms; (2) The more robots are added to the system, the more significant the message transportation (from 67.62\% up to 86.97\% of total processing time in the case of 1 and 150 robots, respectively).

We also evaluate the overhead of containerization microservices using Docker. In this evaluation, we focus on the deployment time and the size of container, as they are indicators for the portability of microservices. We compare the size and build time of bare java file (.jar) with standard docker base image, and a more lightweight docker base image. The results are shown in Fig. 4 and Fig. 5.

\begin{figure}
	\includegraphics[width=\linewidth]{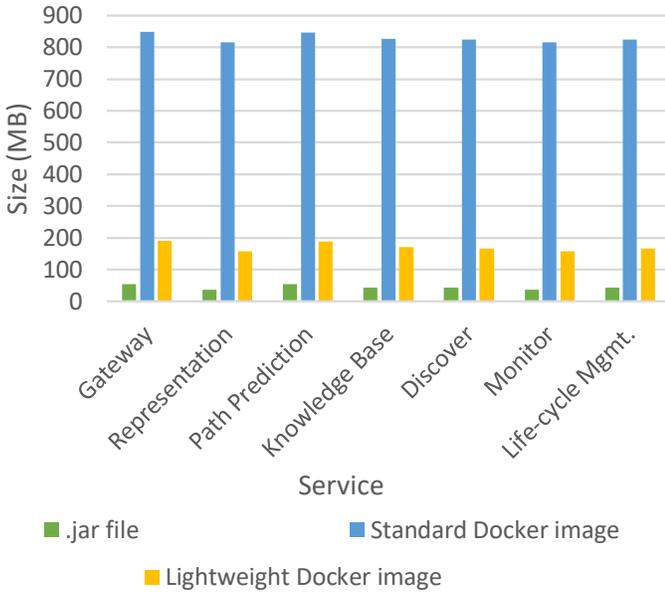}
	\caption{Overhead of containerization regarding the increased size of deployable units.}
	\label{fig:boat5}
\end{figure}

\begin{figure}
	\includegraphics[width=\linewidth]{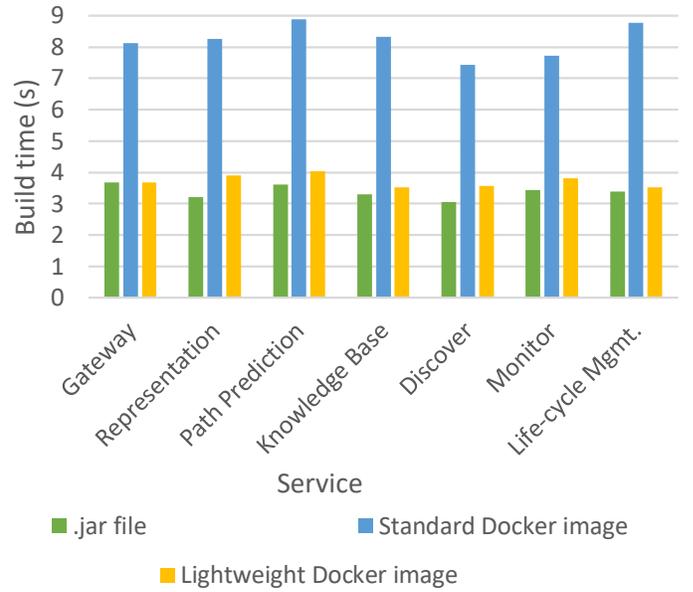}
	\caption{Overhead of containerization regarding the increased build time of deployable units.}
	\label{fig:boat6}
\end{figure}

\subsection{Discussion}

As shown in Fig. 3, when the number of registered robot increases up to 100, the end-to-end processing time increases linearly, but always stays under 20ms. However, beyond this point, the processing time increases significantly but still less than 50ms. In the use case, where the robots move with human walking speed (up to 7km/h), this means the recommendations are communicated back to the fog computing infrastructure within the robot travel distance of 4cm and 10cm, respectively. This result confirms our concern about the latency of message-based communication mechanism implemented in our prototype. When the message queues are filled up with unprocessed messages, the end-to-end delay will increase non-linearly. In a more distributed systems, i.e. microservices are deployed in different physical servers, the significance of message transportation will be even more noticeable. Therefore, the message queues are important indicators  for system's performance and it should be continuously monitored to avoid congestion.

Incorporating container technology not only simplifies the deployment process, but also adds another layer of management and opens up opportunities for advanced features such as service portability. However, it also comes at a cost. As presented in Fig. 4 and Fig. 5, containerizing services increases both the size and the build time of deployable units, which in turn reduces the portability of the service and increases the deployment/redeployment time. Fortunately, to some extent, using lightweight base images can soften the issues.

Refer back to the design criteria we specified in section IV, one of the most noticeable changes in the use of microservices model is that, the application  is decomposed into services. By separating between individual analytics and aggregated analytics as well as dedicating a microservice for a physical entity, the MAIA architecture minimizes the data dependency among microservices by giving each service their own data, which is only accessed by external services via API.

The self-management capability is reflected in the application's ability to monitor itself and adapt to load dynamically during runtime. Although the monitor capability is implemented as an integrated part of the application, it operates independently from other components and works even when other services fail. Running statistics are captured in real-time and used to make scale decisions for services during operation. 

Although the asynchronous messaging pattern poses challenges in guaranteeing the low-latency requirement, there are several possible solutions. Thanks to the portability of microservices in our design, individual analytics can be deployed close to the edge (low latency but limited computing power) or in the cloud (more powerful hardware but also slower). Another strategy to improve the latency is the automatically scale up the services when more requests come, which is done by the \textit{Life-cycle Management} component in our design.

The interoperability of the prototype lies in the use of open standards such as HTTP allowing external entities to communicate with the application. The container technology proves itself to be suitable for microservices, as it provides a straightforward deployment process in different runtime environments. At the same time, \textit{Digital-Physical Integration} is only an abstract layer without any dependency on a concrete technology, allows additional communication technologies to be added. Similarly, new analytics techniques can also be incorporated as new microservices.

The monitor service is able to detect where the failures happen and recover the corresponding microservices. In case of stateful services, the states can be recovered from the persisted data in host servers. In our implementation, this feature is supported by Docker volume. With the use of \textit{Circuit Breaker} pattern, a cascading failure is also prevented.

\section{Conclusion \& Future work}

The primary objective of this paper is to design a distributed architecture for industrial analytics using microservices. Based on the design, we also develop a prototype to evaluate the feasibility of our idea in industrial contexts.

The paper identifies three important trends in industrial analytics: the increase in volume, velocity, and complexity of measured data; the shift of focus from data collection toward data analytics; and the tendency of manufacturers to adopt cloud-based services. These three trends require a new approach for highly scalable and flexible applications. Among others, the microservices architecture promotes the development of applications as a set of independent, autonomous, and self-contained units, which is in line with the trends of future manufacturing. Therefore, we conclude that microservices is a potential candidate for data analytics in industrial contexts. 

The design and implementation of our MAIA architecture underpins both advantages and drawbacks of the microservices architecture. Our evaluation proves that our prototype can achieve an end-to-end latency of less than 20ms for a scenario with up to 100 DTs, and less than 50ms for 150 DTs. The containerization of services does introduce overheads in terms of building time and service size, but using a lightweight base image can help minimizing this burden. Qualitatively saying, the proposed architecture also meets several requirements of decentralization, scalability, fault-tolerance. 


Nevertheless, this paper raised additional questions that need to be addressed. Among others, an effective dynamic service allocation algorithm is required to optimize the system's performance in runtime by balancing between computing power and latency. Also, new messaging techniques should be developed to further improve the latency of microservices.


\section*{Acknowledgment}

This work has received funding by the Federal Ministry for Economic Affairs and Energy (BMWi), Germany under grant no. 01MA17008, project Industrial Communication for Factories (IC4F).



\bibliographystyle{IEEEtran}
\bibliography{bare_conf}
%

\end{document}